\documentclass[aip,jcp,amsmath,amssymb,reprint]{revtex4-1}

\usepackage[utf8]{inputenc}
\usepackage[T1]{fontenc}
\usepackage{graphicx}
\usepackage{dcolumn}
\usepackage{bm}
\usepackage{hyperref}
\usepackage[per-mode=symbol, separate-uncertainty]{siunitx}
\usepackage{cleveref}

\begin{document}
	
	\title{Enhanced spin--orbit coupling in core/shell nanowires} 
	\author{Stephan Furthmeier}
	\author{Florian Dirnberger}
	\affiliation{Institut für Experimentelle und Angewandte Physik, Universität Regensburg, D-93040 Regensburg, Germany}
	\author{Martin Gmitra}
	\affiliation{Institut für Theoretische Physik, Universität Regensburg, D-93040 Regensburg, Germany}
	\author{Andreas Bayer}
	\author{Moritz Forsch}
	\author{Joachim Hubmann}
	\author{Christian Schüller}
	\author{Elisabeth Reiger}
	\affiliation{Institut für Experimentelle und Angewandte Physik, Universität Regensburg, D-93040 Regensburg, Germany}
	\author{Jaroslav Fabian}
	\affiliation{Institut für Theoretische Physik, Universität Regensburg, D-93040 Regensburg, Germany}
	\author{Tobias Korn}
	\author{Dominique Bougeard*}
	\affiliation{Institut für Experimentelle und Angewandte Physik, Universität Regensburg, D-93040 Regensburg, Germany}
	
	
	\begin{abstract}
		The spin--orbit coupling~(SOC) in semiconductors is strongly influenced by structural asymmetries, as prominently observed in bulk crystal structures that lack inversion symmetry. Here, we study an additional effect on the SOC: the asymmetry induced by the large interface area between a nanowire core and its surrounding shell. Our experiments on purely wurtzite GaAs/AlGaAs core/shell nanowires demonstrate optical spin injection into a single free-standing nanowire and determine the effective electron \textit{g}-factor of the hexagonal GaAs wurtzite phase. The spin relaxation is highly anisotropic in time-resolved micro-photoluminescence measurements on single nanowires, showing a significant increase of spin relaxation in external magnetic fields. This behavior is counterintuitive compared to bulk wurtzite crystals. We present a model for the observed electron spin dynamics highlighting the dominant role of the interface-induced SOC in these core/shell nanowires. This enhanced SOC may represent an interesting tuning parameter for the implementation of spin--orbitronic concepts in semiconductor-based structures.\clearpage
	\end{abstract}
	
	\maketitle

	
	The absence of an inversion center in a crystal---which may be inherent to the crystal symmetry or artificially introduced via interfaces, surfaces or doping---induces a splitting of electronic energy bands due to the relativistic effect of spin--orbit coupling~(SOC). This combined interaction of the spin and the orbital degrees of freedom is of fundamental importance for the development of spintronic device concepts, since it has been shown to directly influence the spin dynamics of mobile charge carriers in the crystal, in particular the spin relaxation\cite{Dyakonov1971,Dyakonov1972,Meier1984,Zutic2004,Fabian2007,Dyakonov2008}. The exploration of the physics underlying SOC in condensed matter has quite recently unveiled exciting research areas such as the engineering of SOC in semiconductor heterostructures and hybrid structures, with the prominent example of the realization of a spin helix\cite{Schliemann2003,Bernevig2006,Koralek2009,Walser2012,Sasaki2014,Schoenhuber2014}, spin--orbitronics, i.e. electronic device concepts based on the manipulation of mobile charge and spin carriers\cite{Manchon2015,Dyakonov2008} and, most importantly, the perspective to realize new materials characterized by non-trivial topological order, e.g., topological insulators\cite{Qi2011,Hasan2010,Moore2010} and superconductors\cite{Qi2011,Mourik2012,Das2012}.
	
	Nanowires~(NWs), in particular in non-cen\-tro\-sym\-met\-ric semiconductor III--V materials, offer an ideal testbed to study different microscopic contributions to SOC, given their large surface-to-volume ratio. Combining optimized SOC and the interesting geometrical form factor, NWs are strong candidates for the realization of spin-field-effect transistor concepts\cite{Schliemann2003}, spin--orbit quantum bits\cite{Nadj-Perge2010,Nadj-Perge2012,vandenBerg2013,Frolov2013} or the experimental demonstration of Majorana fermion bound states\cite{Mourik2012,Das2012,Frolov2013}.
	
	Here, we use optical spin orientation, a contact-free and non-invasive method, to study the spin dynamics in undoped and purely wurtzite~(WZ) GaAs/AlGaAs core/shell NWs. Considering a Dyakonov--Perel~(DP) picture of spin relaxation\cite{Dyakonov1971,Dyakonov1972}, we demonstrate the spin--orbit interaction for electrons to be particularly strong in the GaAs NW core. Although our NWs do not show any signs of quantum confinement effects---and can thus be viewed as small pieces of bulk WZ GaAs with a large interface-to-volume ratio---the observed spin dynamics strongly differ from experimental reports for III--V bulk WZ semiconductors such as GaN\cite{Buss2009,Buss2010,Buss2011,Rudolph2014}. We show that the large GaAs/AlGaAs interface area induces a dominant role of interface-related spin--orbit interaction at each NW facet. Our results emphasize the importance of interfaces and their crystallographic orientations in semiconductor heterostructures and hybrid structures when optimizing SOC for spin--orbitronic concepts. 
	
	\section*{Results}
	\subsection*{Optical spin orientation in a single WZ GaAs NW}
	The first objective of our study is to demonstrate efficient optical spin injection into single, free-standing WZ GaAs/AlGaAs core/shell NWs.
	\begin{figure}
		\includegraphics[width=\linewidth]{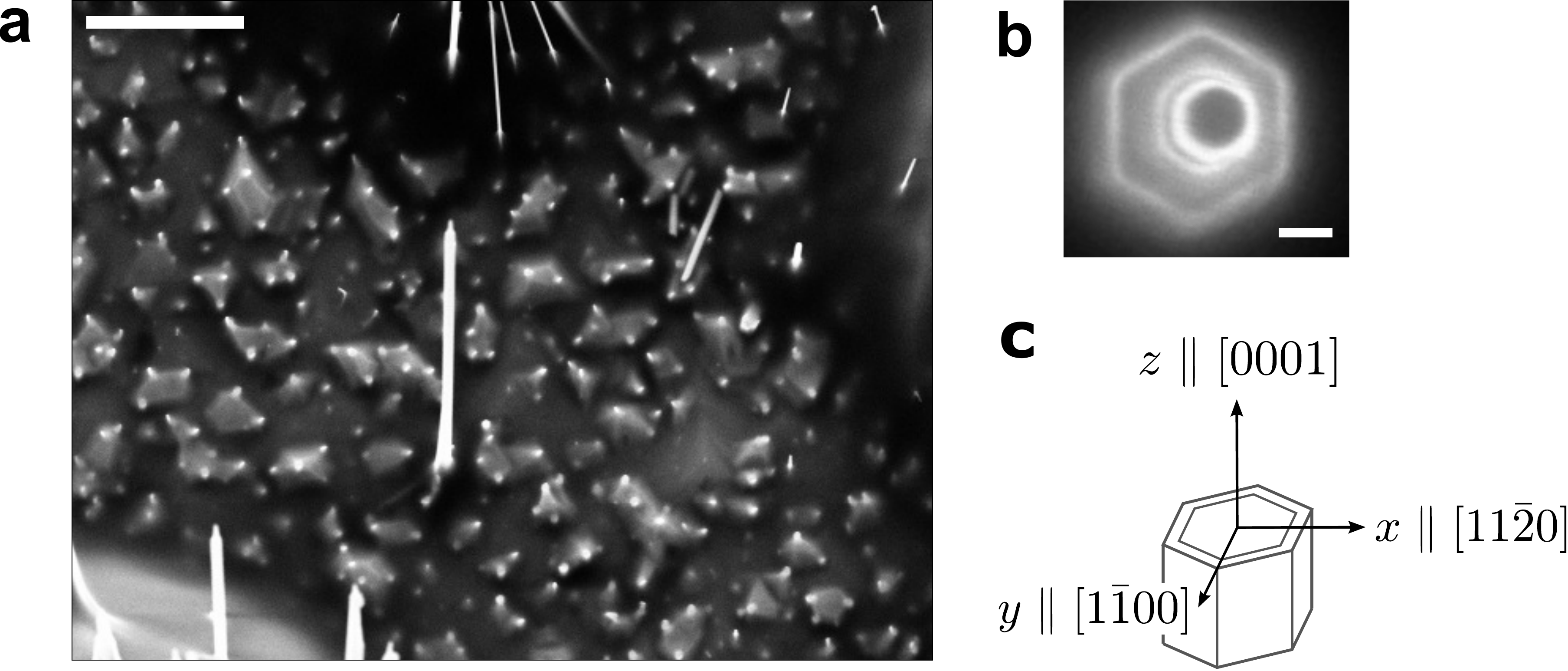}
		\caption{\textbf{Free-standing WZ GaAs/AlGaAs core/shell NWs.} (\textbf{a})~Representative scanning electron microscopic image of a single, free-standing NW used for optical spin orientation experiments. Scale bar, \SI{2}{\um}. (\textbf{b})~Top-view image of the NW in (a), showing the solidified hemispherical catalyst droplet atop the hexagonal NW. Scale bar, \SI{50}{\nm}. (\textbf{c})~Schematic representation of the core/shell NW and the orientation of the coordinate system with respect to its axis, denoted as the $z\parallel[0001]$ axis.\label{fig:SEM}}
	\end{figure}
	Figure~\ref{fig:SEM}a shows a representative scanning electron micrograph of such a wire used for optical orientation measurements. The typical length of these NWs is around \SI{5}{\um} and the diameter \SI{\sim110}{nm}. The crystal structure is purely WZ (see Supplementary Fig.~1). The NWs are nominally intrinsic and grown along the WZ $\hat{c}\parallel\langle0001\rangle$-direction (details can be found in the Methods section)\cite{Furthmeier2014}. Figure~\ref{fig:SEM}b displays the top-view of the same GaAs NW, revealing its characteristic hexagonal cross section and the solidified hemispherical catalyst droplet at the tip. According to transmission electron microscopy, the six equivalent sidewall facets are oriented along the $\langle11\bar{2}0\rangle$-directions of the WZ unit cell, as sketched in Fig.~\ref{fig:SEM}c.
	
Our experiments are based on a confocal approach (see Methods section for details), allowing us to study individual free-standing NWs in polarization-resolved micro-photoluminescence~(\si{\micro}-PL). 
\begin{figure*}
	\centering
	\includegraphics[width=0.75\linewidth]{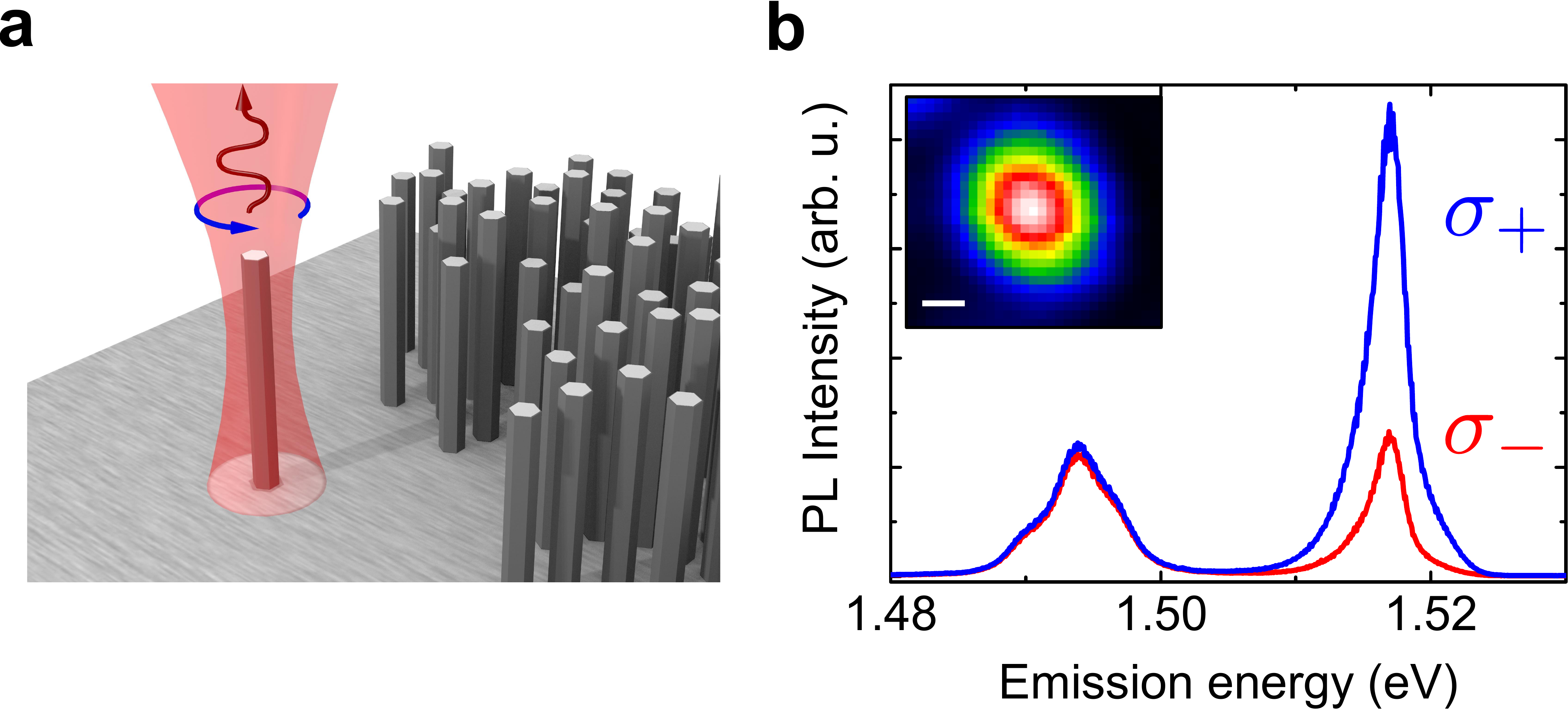}
	\caption{\textbf{Optical spin orientation in single WZ GaAs NWs.} (\textbf{a})~Schematic representation of the measurement geometry used for optical orientation measurements in our experiment. An isolated, free-standing NW is individually excited by a tightly focused circularly polarized laser beam propagating parallel to the NW axis. Helicity-resolved PL emission is detected in confocal configuration along the NW axis. (\textbf{b})~Polarization-dependent \si{\micro}-PL spectra of a single, free-standing WZ GaAs NW at \SI{4.2}{\K} under $\sigma_+$~circularly polarized excitation. While the substrate peak at \SI{\sim1.494}{\eV} holds no significant circular polarization, the NW emission at \SI{1.517}{\eV} is highly polarized with a degree of polarization of \SI{\sim54}{\percent}, indicating efficient optical orientation of spins in the wire. Inset shows an area scan of the integrated PL intensity in the vicinity of this NW in false color coding, evidencing single NW spectroscopy. Scale bar, \SI{1}{\um}.\label{fig:Zero_Field}}	
\end{figure*}
As depicted schematically in Fig.~\ref{fig:Zero_Field}a, each freestanding NW was optically excited with circularly polarized laser light propagating parallel to the NW $\hat{c}\parallel\langle0001\rangle$ axis. We first performed spatially resolved \si{\micro}-PL scans in order to preselect single NWs with highest crystalline purity\cite{Furthmeier2014}. A typical area scan of the integrated PL intensity of such an individual NW is shown in the inset of Fig.~\ref{fig:Zero_Field}b in false color coding, evidencing single NW spectroscopy.
In Fig.~\ref{fig:Zero_Field}b we then show the polarization-resolved \si{\micro}-PL spectra at \SI{4.2}{\K}. Each of them contains two characteristic peaks. The one at $E=\SI{1.494}{\eV}$ is also seen when exciting the bare substrate. It stems from excitons bound to single carbon impurities in the GaAs substrate\cite{Bebb1972,Rao1985}. The luminescence peak at $E=\SI{1.517}{\eV}$ and its narrow linewidth of \SI{3}{\meV} are characteristic for the free-exciton emission in stacking-fault-free WZ GaAs NWs\cite{Ahtapodov2012,Furthmeier2014}.
The polarization-resolved experiment is conducted in the absence of an external magnetic field, under $\sigma_+$~circularly polarized excitation with a near-resonant excitation energy of \SI{1.58}{\eV}, which is estimated to solely induce the heavy hole states-to-conduction band transition in the WZ GaAs core (see Methods section)\cite{De2010,Murayama1994,Ketterer2011-2,Kusch2012,Kim2013,Signorello2014}. The blue curve in Fig.~\ref{fig:Zero_Field}b shows the spectrum obtained for detection of the $\sigma_+$~component of the resulting emission, while the red curve shows the spectrum of the $\sigma_-$~component. The integrated PL intensity $I_+$ of the $\sigma_+$~component of the WZ free exciton is much stronger than its counter-polarized component~$I_-$, revealing significant circular polarization of the luminescence of the WZ GaAs NW in the absence of an external magnetic field. The corresponding degree of circular polarization, defined as $P_\text{C}=\left(I_+-I_-\right)/\left(I_++I_-\right)$, reaches \SI{\sim54}{\percent}. Note that, in contrast, the substrate-related peak shows no significant circular polarization. The large degree of circular polarization of the characteristic NW emission is a strong indication of efficient optical injection of spins into the wire. An optically injected spin ensemble should be depolarized by spin precession when applying an external transverse magnetic field (Hanle effect\cite{Hanle1924}), according to\cite{Meier1984}
\begin{equation}
P_\text{C}(B) = \frac{P_\text{C}(0)}{1+(\omega_\text{L}\tau^*)^2}\,, \qquad \frac{1}{\tau^*} = \frac{1}{\tau_\text{r}}+\frac{1}{\tau_\text{s}}\,, \label{eq:Hanle}
\end{equation}
where $P_\text{C}(0)$ is the degree of polarization at zero magnetic field and $\tau^*$ is the effective spin lifetime, which is given by the inverse sum of the electron lifetime $\tau_\text{r}$ and the spin relaxation time $\tau_\text{s}$. $\omega_\text{L}=g^*\mu_\text{B}B/\hbar$ is the Larmor spin precession frequency induced by the external magnetic field~$(B)$, $\mu_\text{B}$ is the Bohr magneton, $g^*$ is the transverse effective electron $g$-factor, and $\hbar$ is the reduced Planck constant. 
\begin{figure}
	\centering
	\includegraphics[width=0.8\linewidth]{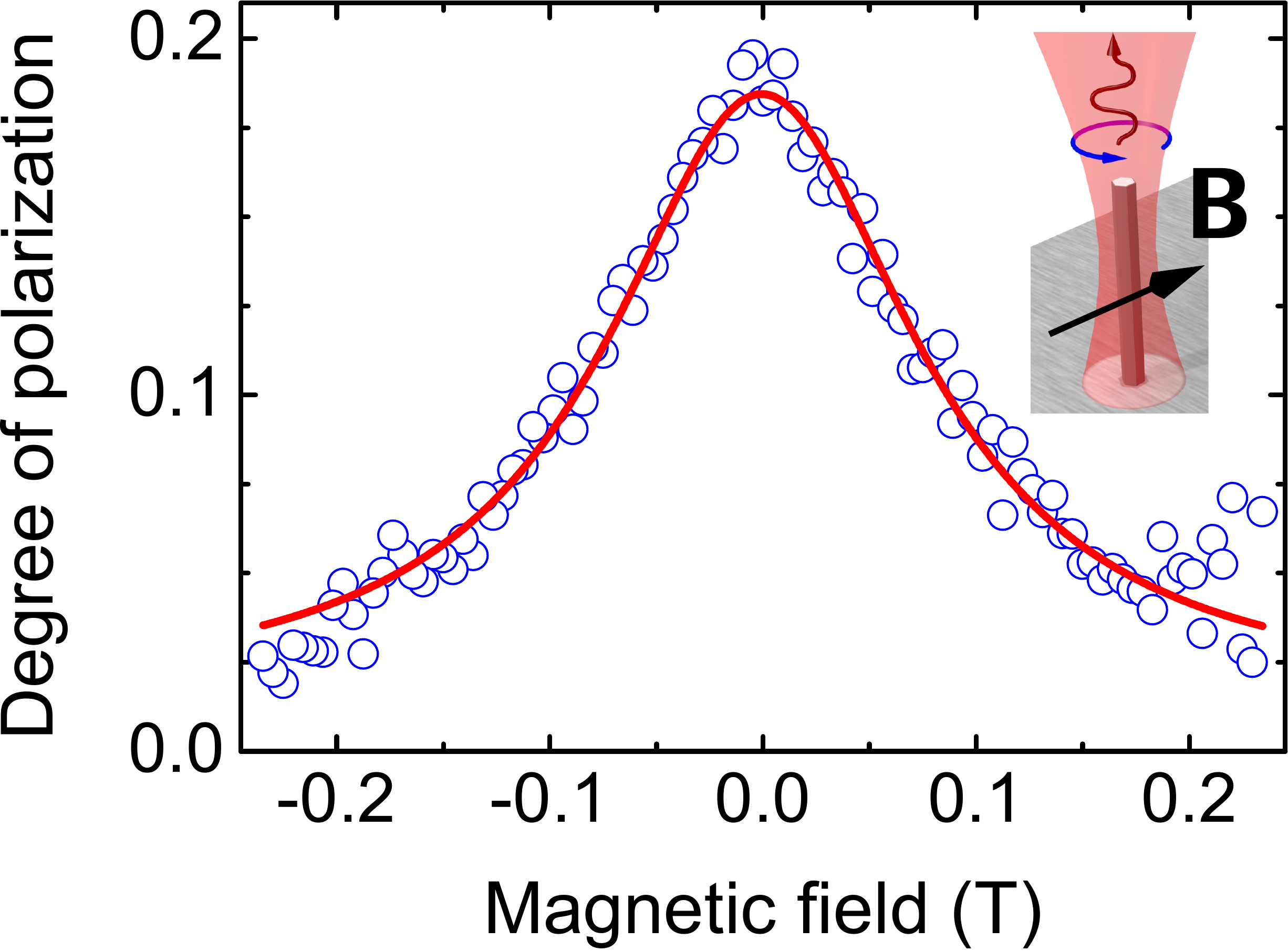}
	\caption{\textbf{Hanle effect measurement for a single WZ GaAs NW.} The degree of polarization of the WZ free exciton emission decreases as a function of the transverse magnetic field. The continuous red line is a Lorentzian fit to the data using equation~(\ref{eq:Hanle}). The schematic measurement configuration is superimposed.\label{fig:Hanle}}
\end{figure}
In Fig.~\ref{fig:Hanle} we thus plot the experimentally determined degree of circular polarization of the WZ free exciton as a function of the magnitude of external magnetic field applied perpendicularly to the NW axis. In addition, the red curve represents a fit of the Hanle function to our data. The excellent agreement of data and fit represents clear evidence for the successful optical injection of a spin ensemble into the single NW. However, without knowing the $g$-factor of WZ GaAs, the time-integrated Hanle measurements cannot be used for quantitative determination of relaxation times of the spin-polarized electron ensemble in the NW.

\subsection*{Spin dynamics in WZ GaAs NWs}
\begin{figure*}
	\includegraphics[width=0.95\linewidth]{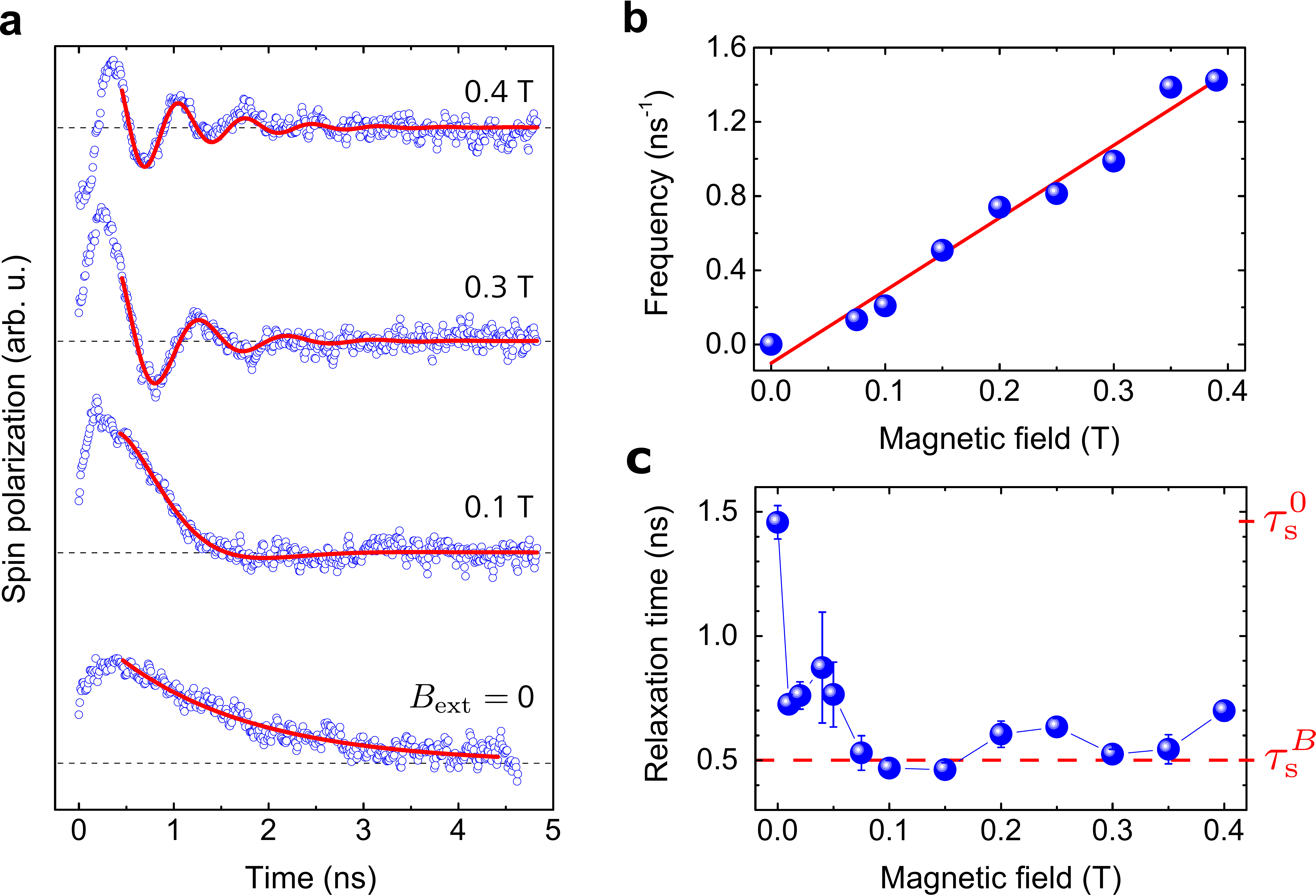}
	\caption{\textbf{Magnetic field effects on the spin dynamics.} (\textbf{a})~Typical time-resolved transients of the spin polarization for a single, free-standing WZ GaAs NW at \SI{4.2}{\K} for transverse applied magnetic fields from $0$ to \SI{400}{\milli\tesla}. The transients are fitted to a single exponential for $B_\text{ext}=0$ and to a damped cosine (continuous red lines) for $B_\text{ext}>0$, respectively, accounting for Larmor precession in non-zero magnetic fields. (\textbf{b})~Extracted values of the Larmor precession frequency as a function of the externally applied transverse magnetic field. From a linear fit to the data (continuous red line) we determine the absolute value of the effective electron $g$-factor in WZ GaAs NWs to be $|g^*|=0.28\pm0.02$. (\textbf{c})~Variation of the spin relaxation time $\tau_\text{s}$ as a function of the transverse applied magnetic field. The sudden decrease of $\tau_\text{s}$ in an external magnetic field reflects an intrinsic spin relaxation anisotropy, with spins pointing along the NW axis relaxing substantially slower than spins perpendicular to the NW axis. The dashed red line marks the drop from the zero-field value $\tau_\text{s}^0$ to $\tau_\text{s}^B\approx{\tau_\text{s}^0}/{3}$ in a transverse field. Error bars correspond to the standard error for fitting the spin polarization transients.\label{fig:Spin_Dynamics}}
\end{figure*}
The dynamics of such a spin ensemble can be accessed in time-resolved \si{\micro}-PL~(TRPL) experiments. The time decay of the \si{\micro}-PL peak of a representative nanowire is shown in Supplementary Fig.~2. We find a long free exciton recombination lifetime of \SI{9.4}{\ns}. Figure~\ref{fig:Spin_Dynamics}a depicts typical time-resolved spin polarization transients of a single, free-standing WZ GaAs NW for externally applied transverse magnetic fields $B_\text{ext}$ from 0 up to \SI{400}{\milli\tesla}.
While for $B_\text{ext}=0$ the decay is characterized by a single exponential, two main features arise as soon as an external magnetic field is applied perpendicular to the NW axis. First, we observe a characteristic oscillatory behavior with increasing frequency for increasing $B_\text{ext}$, and second, a significantly steeper slope of the envelope compared to the zero field transient.

The oscillations in the spin polarization transients arise from spins precessing around the external magnetic field $\mathbf{B}_\text{ext}$ with a frequency corresponding to the Larmor frequency $\omega_\text{L}=g^*\mu_\text{B}B_\text{ext}/\hbar$. According to the optical selection rules, the precession around $\mathbf{B}_\text{ext}$ leads to a periodic change between $\sigma_+$ and $\sigma_-$~polarized luminescence and consequently to the oscillations in the spin polarization transients. The corresponding values of $\omega_\text{L}$ are extracted from the fits and plotted in Fig.~\ref{fig:Spin_Dynamics}b as a function of the applied field $B_\text{ext}$. From a linear fit to the data we calculate the effective $g$-factor in the hexagonal WZ crystal phase of GaAs to be $|g^*|=0.28\pm0.02$. This value is remarkably different from its zincblende counterpart (i.e., $|g^*|=0.44$)\cite{Weisbuch1977}.

\subsection*{Unusual spin relaxation in WZ GaAs NWs}	
The second observation we make is a significantly faster decay of the spin polarization transients for $B_\text{ext}>0$. Since the temporal decay of the spin polarization transients is directly linked to electron spin relaxation, the considerably slower decay of the zero field trace as compared to the envelope for $B_\text{ext}>0$ reflects a distinct increase of spin relaxation in the presence of a transverse magnetic field. To quantify this effect, we determined the corresponding spin relaxation times $\tau_\text{s}^0$~($B_\text{ext}=0$) and $\tau_\text{s}^B$~($B_\text{ext}>0$) by fitting the polarization transients to a monoexponential decay $\exp(-t/\tau_\text{s}^0)$ for zero magnetic field and to a damped cosine $\exp(-t/\tau_\text{s}^B) \cos\left(\omega_\text{L}t\right)$ for $B_\text{ext}>0$, respectively, for several NWs. Figure~\ref{fig:Spin_Dynamics}c exemplarily presents the magnetic field dependence of $\tau_\text{s}$ for a single, free-standing NW in transverse magnetic fields up to \SI{400}{\milli\tesla}. All measured NWs qualitatively show the same behavior: the initially long spin relaxation time $\tau_\text{s}^0\approx\SI{1.5}{\ns}$ drops to a substantially reduced value $\tau_\text{s}^B\approx\SI{0.5}{\ns}$ in the presence of a transverse magnetic field, analogously to the anisotropic spin dephasing found in $(110)$ GaAs quantum wells\cite{Doehrmann2004}. The observed reduction of the spin relaxation time in an external field in Fig.~\ref{fig:Spin_Dynamics}c therefore reflects an intrinsic spin relaxation anisotropy in WZ GaAs NWs: spins pointing along the NW axis ($\parallel$ WZ $\hat{c}$-direction) relax remarkably slower than spins perpendicular to the $\hat{c}$-axis. 

This peculiar magnetic field dependence is counterintuitive when compared to previously reported experiments on spin dynamics in related bulk WZ GaN structures\cite{Buss2009,Buss2010,Buss2011,Rudolph2014}. In these bulk GaN samples, using a similar measurement configuration, an increase of the spin relaxation time is observed when a transverse magnetic field is applied. In order to resolve this puzzle, we present in the next section a model for the anomalous NW spin dynamics developed in the framework of DP spin relaxation\cite{Dyakonov1971,Dyakonov1972}, involving the interface contributions to spin--orbit coupling at the NW sidewall facets.

\section*{Discussion}
Implying DP spin relaxation, which dominates the spin dephasing of free, delocalized electrons in most III--V semiconductor bulk and nano--heterostructure sam\-ples\cite{Zutic2004,Fabian2007}, the effect of SOC on the relaxation time $\tau_\text{s}$ for a given spin component can be described by\cite{Dyakonov1972,Meier1984,Zutic2004,Fabian2007,Dyakonov2008}
\begin{equation}
\frac{1}{\tau_\text{s}} \sim \langle\mathbf{\Omega}_{\mathbf{k},\perp}^2\rangle\,\tau_\text{p}^*\,,
\label{eq:DP_spin_relax_time}
\end{equation}
where $\langle\mathbf{\Omega}_{\mathbf{k},\perp}^2\rangle$ is the mean square effective magnetic field in the plane perpendicular to the considered spin direction and $\tau_\text{p}^*$ is the momentum relaxation time for an individual electron.
As can be seen from equation~(\ref{eq:DP_spin_relax_time}), the DP spin relaxation time sensitively depends on the explicit form of the effective, $\mathbf{k}$-dependent magnetic field $\mathbf{\Omega}_\mathbf{k}$.
Since the diameter of the NW core is large enough to exclude the effects of electron confinement, the inversion asymmetry of the bulk WZ crystal structure determines the form of the intrinsic effective magnetic field\cite{Dresselhaus1955,Rashba1960,Bychkov1984,Margulis1984,LeeYanVoon1996,Lo2005,Wang2007,Fu2008,Buss2009,Buss2010,Rudolph2014}
\begin{equation}
\mathbf{\Omega}_{\mathbf{k}}^\text{bulk} = \begin{pmatrix} \Omega_{\mathbf{k},x}^\text{bulk} \\ \Omega_{\mathbf{k},y}^\text{bulk} \\ \Omega_{\mathbf{k},z}^\text{bulk} \end{pmatrix} = \beta \begin{pmatrix} k_y \\ -k_x \\ 0 \end{pmatrix}\,,    
\label{eq:Omega_bulk}
\end{equation}
where $z\parallel[0001]$ ($\hat{c}$-axis), $x\parallel[11\bar{2}0]$ and $y\parallel[1\bar{1}00]$ (cf. Fig.~\ref{fig:SEM}c). The coefficient $\beta$ is an effective SOC parameter describing the total magnitude of the SO field in bulk WZ (see Supplementary Note~1).
\begin{figure*}
	\includegraphics[width=0.7\linewidth]{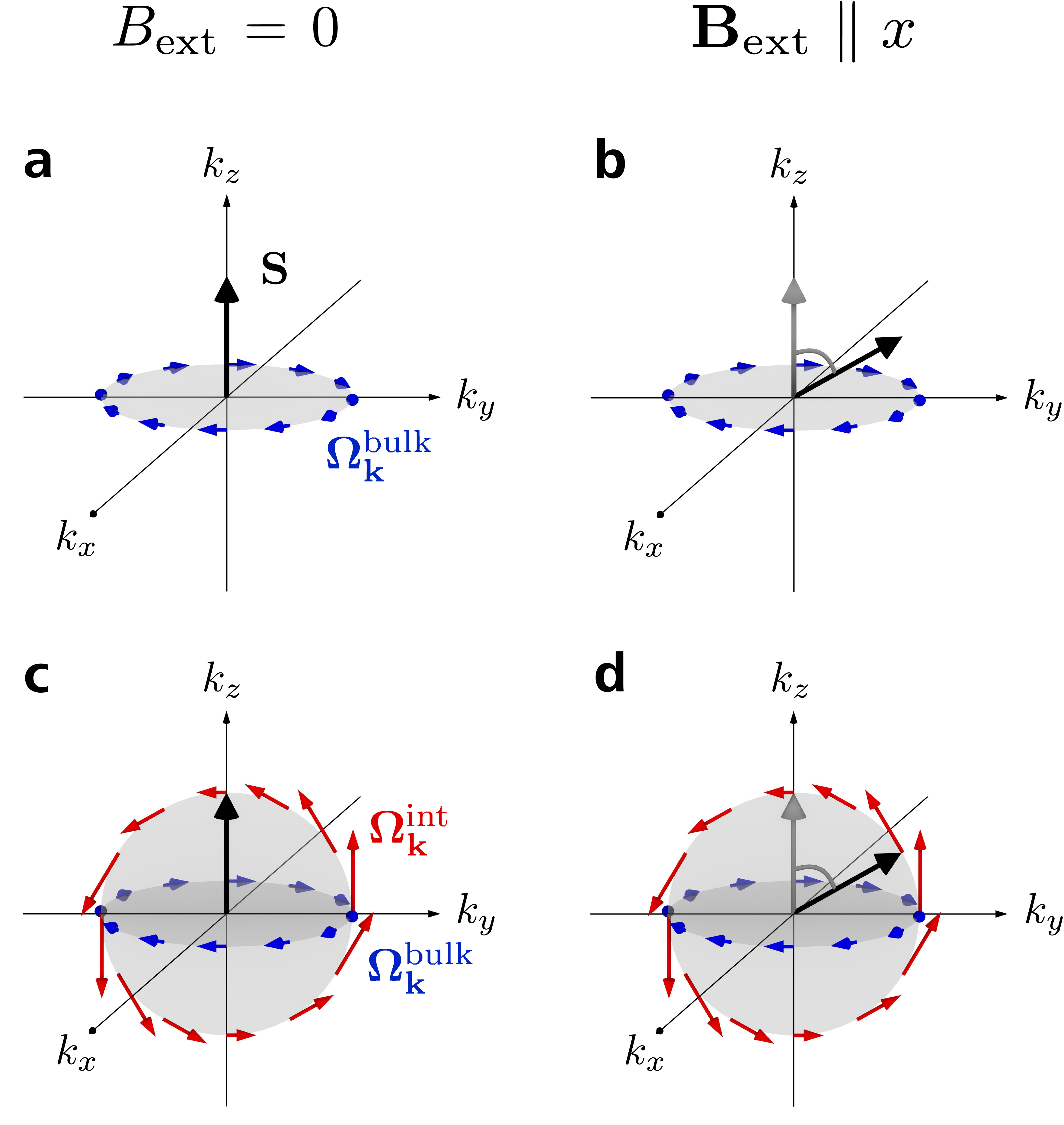}
	\caption{\textbf{Interface-induced SOC effects on the spin relaxation in WZ GaAs/AlGaAs core/shell NWs.} (\textbf{a,b})~Schematic explanation of the spin relaxation anisotropy in bulk WZ crystals: (\textbf{a})~For $B_\text{ext}=0$ the optically generated spin ensemble $\mathbf{S}$ pointing along the $\hat{c}$-axis is susceptible to both the $x$- and $y$-component of the SOC field $\mathbf{\Omega}_{\mathbf{k}}^\text{bulk}$, while (\textbf{b})~Larmor precession in an external magnetic field $\mathbf{B}_\text{ext}\parallel x$ leads to a rotation of $\mathbf{S}$ towards $k_y$, which is then only subject to the $x$-component. (\textbf{c,d})~Schematic model for the observed spin relaxation in WZ GaAs/AlGaAs core/shell NWs, involving interface-induced SOC at the $(11\bar{2}0)$ sidewall facet: (\textbf{c})~For $B_\text{ext}=0$ the optically generated spin ensemble $\mathbf{S}$ pointing along the $\hat{c}$-axis is not susceptible to the largest component $\Omega_{\mathbf{k},z}^\text{int}$ of the SOC fields, while (\textbf{d})~as soon as an external magnetic field $\mathbf{B}_\text{ext}\parallel x$ induces spin precession, this large $z$-component of $\mathbf{\Omega}_{\mathbf{k}}^\text{int}$ starts to act on $\mathbf{S}$ and will quickly dominate the relaxation.\label{fig:Model}}
\end{figure*}
The orientation of $\mathbf{\Omega}_{\mathbf{k}}^\text{bulk}$ is schematically depicted as blue arrows in Fig.~\ref{fig:Model}. For $B_\text{ext}=0$, in Fig.~\ref{fig:Model}a, the spin ensemble $\mathbf{S}$ is optically generated along $k_z$. Since the intrinsic magnetic field $\mathbf{\Omega}_{\mathbf{k}}^\text{bulk}$ which induces a relaxation of $\mathbf{S}$ contains no $k_z$-component, the relaxation is maximized. However, if an external magnetic field is applied transversely to the initial $z$--orientation of the spin ensemble, e.g. $\mathbf{B}_\text{ext}\parallel x$, Larmor precession induced by this external field leads to a rotation of the spin ensemble into the $(k_y,k_z)$--plane, as sketched in Fig.~\ref{fig:Model}b. Then, $\mathbf{S}$ contains only $k_y$- and $k_z$-components. Since $\Omega_{\mathbf{k},y}^\text{bulk}$ cannot act on the $k_y$-component of the spin ensemble and, furthermore, $\Omega_{\mathbf{k},z}^\text{bulk}=0$, the DP mechanism will be less efficient than for the initial situation with $B_\text{ext}=0$. As a consequence, the spin relaxation time $\tau_\text{s}^B$ is expected to be longer than $\tau_\text{s}^0$, which has been confirmed experimentally in bulk WZ GaN samples\cite{Buss2009,Buss2010,Buss2011,Rudolph2014}. The bulk SOC leads to a magnetic field dependence of the spin relaxation time that is opposite to the results from the NWs and is therefore not capable of explaining the observed spin relaxation in our NWs.

Note, however, that in addition to the intrinsic SO field $\mathbf{\Omega}_{\mathbf{k}}^\text{bulk}$, there also exists a contribution to the SOC resulting from the asymmetry of the GaAs/AlGaAs core/shell interface, which is determined by the presence of different atoms at each side of the heterointerface and the corresponding band discontinuities\cite{Lassnig1985,Aleiner1992,Jusserand1995,Ivchenko1996,Pfeffer1999,Roessler2002,Ivchenko2005,Fabian2007,Koralek2009,Devizorova2013,Devizorova2014,Zhou2015}. Although often neglected in bulk and symmetric quantum well samples, these interfacial contributions can be on the same scale as those associated with bulk and structure inversion asymmetry, as demonstrated by recent calculations of the SOC parameters of electrons at a single, atomically sharp GaAs/AlGaAs heterointerface\cite{Devizorova2013,Devizorova2014}. Hexagonal core/shell NWs provide six of these interfaces and a very high ratio of interface area to volume. Consequently, these interfacial SOC effects will be particularly large in such NWs, provided that the electron phase coherence length is smaller than the NW diameter, a situation frequently encountered in vapor--liquid--solid grown NWs. Since we do not observe any signatures of spatial quantum confinement, this latter condition is met in our NWs.

The appropriate form of the interface-induced effective magnetic field, $\mathbf{\Omega}_{\mathbf{k}}^\text{int}$, depends on the crystallographic orientation at the GaAs/AlGaAs core/shell heterointerfaces. Our WZ GaAs NWs exhibit a typical hexagonal cross section with six equivalent $\{11\bar{2}0\}$ sidewall facets, as illustrated in Fig.~\ref{fig:SEM}c. From a symmetry analysis at the respective core/shell interfaces at these facets, we derive that $\mathbf{\Omega}_{\mathbf{k}}^\text{int}$ will always lie in the plane of the core/shell interface and is perpendicular to $\mathbf{k}$. In addition, the SOC arising from one interface is of the $k$-linear Rashba-type. We find this relationship to be equivalent for all six NW sidewall facets. We thus exemplarily discuss the impact of $\mathbf{\Omega}_{\mathbf{k}}^\text{int}$ at facet $(11\bar{2}0)$ oriented along the $x$-direction, a complete evaluation for all facets is given in the Supplementary Note~2. At this $(11\bar{2}0)$ facet, we obtain $\mathbf{\Omega}_{\mathbf{k}}^\text{int}=(0, -\alpha_{\perp}k_z,\alpha_{\parallel}k_y)$. Taking this additional contribution into account, we modify the total effective magnetic field induced by SOC to the expression
\begin{equation}                                                                                                                \mathbf{\Omega}_{\mathbf{k}}=\mathbf{\Omega}_{\mathbf{k}}^\text{bulk}+\mathbf{\Omega}_{\mathbf{k}}^\text{int} = \beta\, \begin{pmatrix} k_y \\ -k_x \\ 0 \end{pmatrix}+\begin{pmatrix} 0 \\ -\alpha_{\perp}k_z \\ \alpha_{\parallel}k_y \end{pmatrix}\,,     
\label{eq:Omega_total}
\end{equation}
where the coefficients $\alpha_{\parallel}$ and $\alpha_{\perp}$ are effective SOC parameters determining the strength of the interfacial contribution parallel and perpendicular to the WZ $\hat{c}$-axis, respectively. Remarkably, due to the low symmetry $\{11\bar{2}0\}$ NW sidewall facets of the $C_\text{s}$ point group, $\alpha_{\parallel}$ and $\alpha_{\perp}$ are linearly independent\cite{Cartoixa2006,Tarasenko2009}, while the size of the bulk contribution is given by the single parameter $\beta$.

The additive action of both SOC-induced fields, $\mathbf{\Omega}_{\mathbf{k}}^\text{bulk}$ and $\mathbf{\Omega}_{\mathbf{k}}^\text{int}$, is illustrated in Fig.~\ref{fig:Model}c and d. An important observation is that the bulk effective field (blue arrows as in Fig.~\ref{fig:Model}a, b) lies in the $(k_x,k_y)$--plane, while the effective field resulting from the heterointerface (red arrows) lies in the $(k_y,k_z)$--plane, for our example of the facet $(11\bar{2}0)$. Thus, compared to the pure bulk situation sketched in Fig.~\ref{fig:Model}a and b, in the NWs the core/shell interface obviously introduces a non-zero $k_z$--component to the total effective magnetic field. This particular component may now alter the relaxation time of the spin ensemble $\mathbf{S}$. In addition, when $\alpha_{\parallel}$ and $\alpha_{\perp}$ are different, the magnitude of the effective magnetic field components due to interface inversion asymmetry differs for $k_y$ and $k_z$. This is illustrated through different vector norms of the $\mathbf{\Omega}_{\mathbf{k}}^\text{int}$ components in Fig.~\ref{fig:Model}c and d, in contrast to the constant vector norms of the contributions from $\mathbf{\Omega}_{\mathbf{k}}^\text{bulk}$.

Let us first consider the case $\alpha_{\parallel}>\alpha_{\perp}$ and $\alpha_{\parallel}>\beta$. As a consequence, the magnitude of the interface-induced effective magnetic field is stronger in $k_z$-- than in $k_y$--direction. It is also larger than the magnitude of the effective magnetic field due to SOC in bulk. This situation is sketched in Fig.~\ref{fig:Model}c and d. For $B_\text{ext}=0$, the spin ensemble $\mathbf{S}$ is optically generated along $k_z$. Thus, the largest component of $\mathbf{\Omega}_{\mathbf{k}}$ is parallel to $\mathbf{S}$, as shown in Fig.~\ref{fig:Model}c, and cannot contribute to the spin relaxation. However, as soon as an external magnetic field induces a precession of the spin ensemble, this large component of $\mathbf{\Omega}_{\mathbf{k}}$ starts to act on $\mathbf{S}$, as illustrated in Fig.~\ref{fig:Model}d, and will quickly dominate the relaxation. As a consequence, the spin relaxation time $\tau_\text{s}^B$ will then be shorter than $\tau_\text{s}^0$. This precisely describes our experimental findings discussed in Fig.~\ref{fig:Spin_Dynamics}c. If however, we consider all the other possible relations of $\alpha_{\parallel}, \alpha_{\perp}$ and $\beta$, they impose $\tau_\text{s}^B>\tau_\text{s}^0$. This relation is observed in the bulk WZ material \cite{Buss2009,Buss2010,Buss2011,Rudolph2014} and is opposite to the results of our study.

Our model thus not only supplies a picture for the experimental observations, but also provides information on the relationships of the three effective SOC parameters. Taking into account the complete evaluation for all NW sidewall facets developed in the Supplementary Note~2 we find
\begin{equation}
\frac{\alpha^{2}_{\parallel}}{2\beta^{2}+\alpha^{2}_{\perp}} \approx 4\,,
\label{eq:SOC_ratio}
\end{equation}
which implies that the interface-induced $z$-contribution to the effective SO field $\mathbf{\Omega}_\mathbf{k}$ is significantly larger than the $x$--$y$-contributions from both bulk ($\alpha_{\parallel}>\beta$) and the interfaces ($\alpha_{\parallel}>\alpha_{\perp}$).

In conclusion, our study on GaAs NWs provides insight both into the specific WZ GaAs/AlGaAs system and, more fundamentally, into the SOC of a core/shell NW. To create a spin ensemble, we have demonstrated efficient optical spin injection into single free-standing NWs. We determined the effective $g$-factor in the hexagonal WZ phase of GaAs to be $|g^*|=0.28\pm0.02$ and obtained long spin relaxation times up to \SI{\sim1.5}{\ns} in our WZ GaAs NWs. A more fundamental implication emerges from the peculiar dynamics of spin relaxation in the NW core/shell structure. Our study suggests that a dominant contribution to SOC originates from the NW core/shell interface, which, by its nature, breaks the symmetry of the bulk lattice. For the common situation of the transport phase coherence length being shorter than the diameter of the NW, the presence of the interface will strongly modify the SOC. We believe that this effect plays an important role in experiments, where high SOC in a semiconductor NW is required. Beyond that, it could offer an opportunity to tune the SOC in semiconductor heterostructures and hybrid structures by choosing proper crystallographic orientations and material compositions of heterointerfaces.

\section*{Methods}
\subsection*{NW growth and characterization}
The investigated GaAs/AlGaAs core/shell NWs were synthesized by molecular beam epitaxy (MBE) using the Au-assisted vapor--liquid--solid growth mechanism\cite{Wagner1964,Messing2009} on GaAs$\left(111\right)_{\textrm{B}}$ substrates covered with \SI{10}{\angstrom}~Au. By tuning of MBE growth parameters, the crystalline structure was optimized to produce nominally undoped NWs with a defect-free, pure WZ GaAs phase over lengths of several \si{\um} (see Supplementary Fig.~1). Growth for 125~min using a Ga deposition rate of \SI{0.8}{\angstrom\per\second} and an adjusted As$_{4}$ beam equivalent pressure of \SI{3e-6}{Torr} resulted in NWs grown along the WZ $\hat{c}$-axis with a length of \SI{\sim5}{\um} and a core diameter of \SI{\sim80}{\nm}. In order to disable the dominant nonradiative recombination at the bare GaAs surface\cite{Demichel2010,Chang2012}, the NWs were passivated with a uniform $\textrm{Al}_{0.36}\textrm{Ga}_{0.64}\textrm{As}$ shell with a thickness of \SI{\sim10}{\nm}, ensuring intense and robust PL of the GaAs core. This shell was surrounded by a \SI{\sim5}{\nm} GaAs cap, which protects the $\textrm{Al}_{0.36}\textrm{Ga}_{0.64}\textrm{As}$ layer from oxidation\cite{Perera2008} and allows for optical measurements of the same NWs over weeks without any deterioration. Combined \si{\micro}-PL and transmission electron microscopic studies performed on single wires showed that the NWs produced provide a pure WZ crystal structure and are of very high crystalline and optical quality\cite{Furthmeier2014}.

\subsection*{Optical spin orientation in single NWs}
After growth, the NW areal density was initially reduced by an ultrasonic bath dip and then patterns were created on the sample, which permits us to identify and optically study individual free-standing NWs. We first performed spatially resolved \si{\micro}-PL area scans in order to preselect single NWs with highest crystalline purity (Fig.~\ref{fig:Zero_Field}b). The NWs of interest were then individually further investigated by polarization- and time-resolved \si{\micro}-PL before characterizing their structural properties with scanning electron microscopy. The PL and TRPL measurements were conducted at \SI{4.2}{\K} in a confocal configuration, where both the excitation and collection beams were  parallel to the sample normal and along the NW axes (Fig.~\ref{fig:Zero_Field}a). Single free-standing NWs were excited with the \SI{785}{\nm} emission line~(\SI{\sim1.58}{\eV}) of a semiconductor laser diode operating in continuous-wave or pulsed mode (\SI{70}{\ps} pulses at a repetition rate of \SI{1}{\MHz}), which was chosen to be near-resonant in order to solely induce the heavy hole states-to-conduction band transition in the WZ GaAs core\cite{De2010,Murayama1994,Ketterer2011-2,Kusch2012,Kim2013,Signorello2014}. The laser light was focused down to the sample using a $100\times$ microscope objective with $\textrm{NA}=0.8$ that provides a minimum spot diameter of around \SI{1}{\um}. The emitted PL was collected along the NW axis by the same objective and imaged onto the entrance slit of a grating spectrometer. Time-integrated \si{\micro}-PL spectra were acquired using a cooled charge coupled device, while TRPL signals were detected by a Hamamatsu streak camera system with a time resolution of \SI{\sim50}{\ps}. For the optical orientation experiments the incident near-resonant excitation was $\sigma_+$~circularly polarized, which, according to the WZ optical interband selection rules\cite{Birman1959-1,Birman1959-2,Tronc1999}, injects spin-polarized carriers in the GaAs NW core where the light is absorbed. These carriers recombine to produce polarized luminescence, which is resolved into right~($I_+$) and left~($I_-$) circularly polarized components, defining the degree of circular polarization: $P_\text{C}=\left(I_+-I_-\right)/\left(I_++I_-\right)$. The Hanle effect and time-resolved \si{\micro}-PL measurements were performed by mounting the sample between the coils of an electromagnet, where magnetic fields up to \SI{400}{\milli\tesla} could be applied in the sample plane perpendicular to the NW axes.

\section*{References}
\bibliographystyle{naturemag}

\section*{Acknowledgements}
We gratefully acknowledge financial support by the German Research Foundation (DFG) via SFB~689.

\end{document}